\documentclass[prd,a4paper,nofootinbib,floatfix,onecollumn,notitlepage,amssymb]{revtex4-1}
\usepackage{amsmath}
\usepackage{graphicx}

\begin{document}

\title{Stability of relativistic Bondi accretion in Schwarzschild--(anti-)de Sitter spacetimes}
\author{Patryk Mach and Edward Malec}
\affiliation{M.~Smoluchowski Institute of Physics, Jagiellonian University, Reymonta 4, 30-059 Krak\'{o}w}

\begin{abstract}
In a recent paper we investigated stationary, relativistic Bondi-type accretion in Schwarzschild--(anti-)de Sitter spacetimes. Here we study their stability, using the method developed by Moncrief. The analysis applies to perturbations satisfying the potential flow condition. We prove that global isothermal flows in Schwarzschild--anti-de Sitter spacetimes are stable, assuming the test-fluid approximation. Isothermal flows in Schwarzschild--de Sitter geometries and polytropic flows in Schwarzschild--de Sitter and Schwarzschild--anti-de Sitter spacetimes can be stable, under suitable boundary conditions.

\end{abstract}

\maketitle

\section{Introduction}

In~\cite{kmm} we have obtained a family of solutions describing spherically symmetric, steady accretion of perfect fluids in Schwarzschild--(anti-)de Sitter spacetimes. Here we study linear stability of these solutions. The proof follows strictly the work of Moncrief~\cite{moncrief}, who showed linear stability of analogous solutions in the Schwarzschild case.

In~\cite{kmm} we dealt with transonic solutions for polytropic, $p = K \rho^\Gamma$, and ``isothermal'' equations of state of the form $p = ke$ with $k = 1/3, 1/2, 1$ (for these values of $k$ one obtains analytic solutions in a closed form). Here $p$ denotes the pressure, $\rho$ the baryonic density, $e$ is the energy density, and $K$ and $\Gamma$ are constants.

In this paper we focus only on those transonic solutions that are subsonic far from the central black hole and supersonic in its vicinity. Usually, there exists another branch of solutions that is subsonic for radii smaller than the sonic radius $r_\ast$, and supersonic for $r > r_\ast$. This branch has a natural interpretation of ``wind'', instead of an accretion flow. We will not discuss the stability of such solutions.

The strategy adopted in~\cite{kmm} was as follows. We fix some, usually finite, radius $r = r_\infty$ of the boundary of the cloud, the boundary value of $\rho$ or $e$ (denoted as $\rho_\infty$ and $e_\infty$), and, in the case of polytropic equations of state, the polytropic exponent $\Gamma$ and the boundary value of the local speed of sound $a_\infty$. For those parameters we search for a transonic solution that can be continued inward, at least up to the horizon of the black hole. The condition that the solution passes through a sonic point fixes the boundary value of the radial velocity, but one has to verify that the flow is subsonic in the outer part of the accretion cloud.

For a negative cosmological constant $\Lambda$ and the assumed ``isothermal'' equations of state such solutions can be also continued outward, up to infinite radii, where the energy density tends to zero (note that in the Schwarzschild spacetime $e \to e_\infty \neq 0$ as $r \to \infty$~\cite{michel}). This is not possible for $\Lambda > 0$, in which case the solutions obtained for the ``isothermal'' equations of state diverge at the cosmological horizon. Polytropic solutions can only be continued to arbitrarily large radii for $\Lambda = 0$.

\section{Notation}

In this paper we follow notation conventions of~\cite{kmm}. The flow satisfies the continuity equation
\begin{equation}
\label{aaa}
\nabla_\mu \left( \rho u^\mu \right) = 0
\end{equation}
and the energy--momentum conservation law
\begin{equation}
\label{aab}
\nabla_\mu T^{\mu \nu} \equiv \nabla_\mu \left[ (e + p) u^\mu u^\nu + p g^{\mu \nu} \right] = 0.
\end{equation}
Here, as already stated in the Introduction, $\rho$ denotes the baryonic density; $e$ is the energy density; $p$ is the pressure; $u^\mu$ are components of the four-velocity of the fluid. In the following we will also use the specific enthalpy $h = (e + p)/\rho$. The local speed of sound will be denoted by $a$.

We work in geometrical units with $c = G = 1$ and assume the signature of the metric $g_{\mu\nu}$ of the form $(-,+,+,+)$. Greek indices run through spacetime dimensions $0,1,2,3$. Latin indices are used for spatial dimensions only. The determinant of the metric $g_{\mu\nu}$ will be denoted by $\mathrm{det} g$. The projection tensor is defined as
\[ P_\mu^{\;\; \nu} = \delta_\mu^{\;\; \nu} + u_\mu u^\nu. \]
Throughout this paper we work in spherical coordinates $(t,r,\theta,\phi)$.

\section{Stationary solutions}

We consider Schwarzschild--(anti-)de Sitter spacetimes with the line element
\begin{equation}
\label{aae}
ds^2 = - \left( 1 - \frac{2m}{r} - \frac{\Lambda}{3} r^2 \right) dt^2 + \frac{dr^2}{1 - \frac{2m}{r} - \frac{\Lambda}{3} r^2} + r^2 \left( d\theta^2 + \sin^2 \theta d\phi^2 \right).
\end{equation}

In \cite{kmm} we obtained analytic solutions for equations of state of the form $p = ke$, where $k = 1/3, 1/2$, and 1. The following simple formulae hold in all above cases: $\rho = A/(r^2 u^r)$, $e = B \rho^{1 + k}$, $h = (1 + k) B \rho^k$. Here $A$ and $B$ are constant. The local speed of sound is also constant: $a^2 = k$.

The solution for $k=1/3$ is given by the following formulae. We define
\begin{eqnarray*}
X_1 & = & - 1 + \frac{2m}{r} + \frac{\Lambda}{3}r^2 +  \frac{(1 - 9\Lambda m^2) r^2}{(3 m)^2} \cos \left\{ \frac{\pi}{3} - \frac{1}{3} \mathrm{arc \, cos} \left[ \frac{3^3 m^2 \left( 1 - \frac{2m}{r} - \frac{\Lambda}{3} r^2 \right)}{(1 - 9 \Lambda m^2) r^2} \right] \right\}, \\
X_2 & = & - 1 + \frac{2m}{r} + \frac{\Lambda}{3}r^2 + \frac{(1 - 9\Lambda m^2) r^2}{(3 m)^2} \cos \left\{ \frac{\pi}{3} + \frac{1}{3} \mathrm{arc \, cos} \left[ \frac{3^3 m^2 \left( 1 - \frac{2m}{r} - \frac{\Lambda}{3} r^2 \right)}{(1 - 9 \Lambda m^2) r^2} \right] \right\}. \\
\end{eqnarray*}
Then the square of the radial component of the four-velocity of the solution branch that is subsonic outside the sonic point can be written as
\begin{equation}
\label{zz1}
(u^r)^2 = \left\{\begin{array}{ll} X_2, & r \ge 3m, \\ X_1, & r < 3m. \end{array} \right.
\end{equation}
The sonic point is defined as a location where $a^2 = (u^r/u_t)^2$. For $k = 1/3$ it is located at $r_\ast = 3m$, irrespectively of the value of the cosmological constant.

For $k = 1/2$ the sonic radius is given by
\[ r_\ast = \left\{ \begin{array}{ll}
2 \sqrt{\frac{2}{|\Lambda|}} \sinh \left[ \frac{1}{3} \mathrm{ar\, sinh} \left( \frac{15 m \sqrt{|\Lambda|}}{4 \sqrt{2}} \right) \right], & \Lambda < 0, \\
5m/2, & \Lambda = 0,\\
2 \sqrt{\frac{2}{\Lambda}} \cos \left[ \frac{\pi}{3} + \frac{1}{3} \mathrm{arc \, cos} \left( \frac{15 m \sqrt{\Lambda}}{4 \sqrt{2}} \right) \right], & 0 < \Lambda < 1/(9m^2).
\end{array} \right. \]
The radial component of the velocity can be expressed as
\begin{equation}
\label{zbb}
u^r = \left\{ \begin{array}{ll} B_\ast r^2 + \sqrt{B_\ast^2 r^4 - 1 + \frac{2m}{r} + \frac{\Lambda}{3} r^2}, & r \ge r_\ast, \\ B_\ast r^2 - \sqrt{B_\ast^2 r^4 - 1 + \frac{2m}{r} + \frac{\Lambda}{3} r^2}, & r < r_\ast, \end{array} \right.
\end{equation}
where
\[ B_\ast = - \frac{1}{r_\ast^2} \sqrt{\frac{3m}{r_\ast} - 1}. \]

For $k=1$ one obtains
\begin{equation}
\label{ai}
(u^r)^2 = \frac{1 - \frac{2m}{r} - \frac{\Lambda}{3}r^2}{\left( \frac{r}{r_h} \right)^4 - 1},
\end{equation}
where the areal radius of the black hole horizon is given by
\[ r_h = \left\{ \begin{array}{ll}
\frac{2}{\sqrt{|\lambda|}} \sinh \left[ \frac{1}{3} \mathrm{ar \, sinh} \left( 3 m \sqrt{|\Lambda|} \right) \right], & \Lambda < 0, \\
2m, & \Lambda = 0, \\
\frac{2}{\sqrt{\Lambda}} \cos \left[ \frac{\pi}{3} + \frac{1}{3} \mathrm{arc \, cos} \left( 3 m \sqrt{\Lambda} \right) \right], & 0 < \Lambda < 1/(9m^2).
 \end{array} \right. \]

For $\Lambda > 0$ the above solutions extend only up to the cosmological horizon, i.e., for
\[ r < \frac{2}{\sqrt{\Lambda}} \cos \left[ \frac{\pi}{3} - \frac{1}{3} \mathrm{arc \, cos} \left( 3 m \sqrt{\Lambda} \right) \right] \]
(note that we are only dealing with static Schwarzschild--de Sitter spacetimes with $\Lambda < 1/(9 m^2)$). For $\Lambda < 0$ the solutions cover the entire spacetime (except for the singularity at $r = 0$), with $\rho \to 0$ as $r \to \infty$.

In~\cite{kmm} we also investigated polytropic solutions, both for positive and negative values of $\Lambda$. Unlike ``isothermal'' solutions, they cannot be expressed in a closed form, but they can be easily computed numerically, by solving algebraic equations only. As already mentioned in the Introduction, they can be extended to arbitrarily large radii only for $\Lambda = 0$. The absence of global polytropic solutions is not very surprising for $\Lambda > 0$ because of the existence of cosmological horizons, but it is not obvious for $\Lambda < 0$. There is an interesting mechanism that is responsible for this fact, and it is discussed in~\cite{kmm}. The stability analysis presented below applies also for polytropic solutions, modulo a reservation concerning boundary conditions.

\section{Moncrief's method}

The technique used in~\cite{moncrief} allows for a linear stability analysis against isentropic perturbations for which the vorticity tensor
\[ \omega_{\mu \nu} = P^{\;\; \alpha}_\mu P^{\;\; \beta}_\nu \left[ \nabla_\beta (h u_\alpha) - \nabla_\alpha (h u_\beta) \right] \]
vanishes. We say that such perturbations satisfy the potential flow condition.

It is easy to show that for an isentropic flow, for which $dh = dp/\rho$, Eqs.~(\ref{aaa}) and (\ref{aab}) yield
\[ u^\mu \nabla \left( h u_\nu \right) + \partial_\nu h = 0. \]
Using the above equation, one gets
\[ \omega_{\mu \nu} = \nabla_\nu \left( h u_\mu \right) - \nabla_\mu \left( h u_\nu \right) = \partial_\nu \left( h u_\mu \right) - \partial_\mu \left( h u_\nu \right) . \]
Thus, if $\omega_{\mu\nu} = 0$, the vector field $h u_\mu$ can be expressed (locally) as a gradient of a potential, i.e., $h u_\mu = \partial_\mu \psi$. In this case, the normalization condition for the four-velocity, $u_\mu u^\nu = -1$, yields $h = \sqrt{ - \partial_\mu \psi \partial^\mu \psi}$ and $u_\mu = \partial_\mu \psi / \sqrt{- \partial_\nu \psi \partial^\nu \psi}$. Most notably, the continuity equation (\ref{aaa}) can be simplified to the scalar equation
\begin{equation}
\label{aac}
\nabla_\mu \left( \frac{\rho}{h} \nabla^\mu \psi \right) = 0.
\end{equation}
It can be also shown that stationary, isentropic Bondi-type flows satisfy the potential flow condition, i.e., $\omega_{\mu \nu} = 0$.

The equation governing linear perturbations of the potential $\delta \psi$ can be obtained directly from Eq.~(\ref{aac}). It reads
\begin{equation}
\label{aad}
\frac{1}{\sqrt{\mathrm{- det} \mathfrak G}} \partial_\mu \left( \sqrt{- \mathrm{det} \mathfrak G} \mathfrak G^{\mu\nu} \partial_\nu \delta \psi \right) = 0,
\end{equation}
where $\mathfrak G_{\mu\nu}$ denotes the Lorentzian metric
\[ \mathfrak{G}_{\mu\nu} = \frac{\rho}{ah} \left[ g_{\mu\nu} + \left( 1 - a^2 \right) u_\mu u_\nu \right]. \]
The inverse matrix and the square root of the absolute value of its determinant read
\[ \mathfrak G^{\mu \nu} = \frac{ah}{\rho} \left[ g^{\mu\nu} - \left( \frac{1}{a^2} - 1 \right) u^\mu u^\nu \right], \quad \sqrt{- \mathrm{det} \mathfrak G} = \frac{\rho^2}{a h^2} \sqrt{- \mathrm{det} g}. \]

Equation (\ref{aad}) has a form of a wave equation for $\delta \psi$, with respect to the metric $\mathfrak G_{\mu\nu}$. There is a related energy--momentum tensor
\[ \mathfrak T_\mu^{\;\; \nu} = \frac{1}{2} \left( \mathfrak G^{\nu \alpha} \partial_\mu \delta \psi \partial_\alpha \delta \psi - \frac{1}{2} \delta_\mu^{\;\; \nu} \mathfrak G^{\alpha \beta} \partial_\alpha \delta \psi \partial_\beta \delta \psi \right) \]
and the perturbation energy measure
\[ E = -2 \int_\Omega d^3 x \sqrt{- \mathrm{det} \mathfrak G} \mathfrak T_t^{\;\; t} = \int_\Omega d^3 x \sqrt{- \mathrm{det} \mathfrak G} \left[ - \frac{1}{2} \mathfrak G^{tt} \left( \partial_t \delta \psi \right)^2 + \frac{1}{2} \mathfrak G^{ij} \partial_i \delta \psi \partial_j \delta \psi \right], \]
where the integration is carried out over a space-like, $t = \mathrm{const}$ region $\Omega$. In the following, we are interested in perturbations on the spherically symmetric, steady accretion flow on the background metric given by Eq.~(\ref{aae}). The energy of perturbations comprised in a region between two radii $r_1$ and $r_2$ can be written as
\[ E_{(r_1,r_2)} = \int_{r_1}^{r_2}dr \int_0^\pi d\theta \int_0^{2\pi} d\phi \sqrt{- \mathrm{det} \mathfrak G} \left[  - \frac{1}{2} \mathfrak G^{tt} \left( \partial_t \delta \psi \right)^2 + \frac{1}{2} \mathfrak G^{rr} (\partial_r \delta \psi)^2 \right]. \]
The time derivative of $E_{(r_1,r_2)}$ can be easily computed:
\begin{eqnarray}
\label{bba}
\frac{d}{dt} E_{(r_1,r_2)} & = & 2 \left. \int_0^\pi d\theta \int_0^{2\pi} d\phi \sqrt{- \mathrm{det} \mathfrak G} \mathfrak T_t^{\;\; r} \right\vert_{r_1}^{r2} \\
& = & \left. \int_0^\pi d\theta \int_0^{2\pi} d\phi \sqrt{- \mathrm{det} \mathfrak G} \left[ \mathfrak G^{rt} (\partial_t \delta \psi)^2 + \mathfrak G^{rr} \partial_t \delta \psi \partial_r \delta \psi \right] \right\vert_{r_1}^{r2}. \nonumber
\end{eqnarray}

The strategy of proving the linear stability is standard, but there are certain adjustments, specific to the accretion problem. The first step is to note that it is enough to focus on perturbations that are located outside the sonic point. Intuitively, this is because linear acoustic perturbations cannot escape from the supersonic region with $r < r_\ast$. Formally, the surface $r = r_\ast$ is a null surface with respect to the metric $\mathfrak G_{\mu\nu}$, located outside the event horizon of the black hole~\cite{moncrief}. Let $r_\infty$ denote the radius of the outer boundary of the accretion cloud. As pointed earlier, for $\Lambda < 0$ and ``isothermal'' equations of state discussed in Sec.~2, one can set $r_\infty = \infty$, but in general $r_\infty$ has to be finite. The energy of the perturbations located outside the sonic horizon is equal to $E_{(r_\ast,r_\infty)}$. In the second step of the proof one has to show that $E_{(r_\ast,r_\infty)}$ is positive definite. The last step consists of establishing conditions, under which $dE_{(r_\ast,r_\infty)}/dt \le 0$. We say that the accretion flow is linearly stable, provided that the above conditions are satisfied.

Positivity of $E_{(r_\ast, r_\infty)}$ follows by a direct inspection. Working in coordinates (\ref{aae}), we have
\[ \sqrt{- \mathrm{det} \mathfrak G} \mathfrak G^{tt} = - \sqrt{\mathrm{det} g} \frac{\rho (u_t)^2}{a^2 h} \frac{ 1 - a^2 \left( \frac{u^r}{u_t} \right)^2 }{\left( 1 - \frac{2m}{r} - \frac{\Lambda}{3} r^2 \right)^2}, \quad \sqrt{- \mathrm{det} \mathfrak G} \mathfrak G^{rr} = \sqrt{\mathrm{det} g} \frac{\rho (u_t)^2}{a^2 h} \left[ a^2 - \left( \frac{u^r}{u_t} \right)^2 \right] \]
(remember that $a^2 > (u^r/u_t)^2$ for $r > r_\ast$).

There are two boundary terms in the expression for $dE_{(r_\ast,r_\infty)}/dt$ given by Eq.~(\ref{bba}). The inner boundary term (for $r = r_\ast$) is clearly nonpositive. Note that $\sqrt{-\mathrm{det} \mathfrak G} \mathfrak G^{rr}$ vanishes at $r = r_\ast$, while
\[ \sqrt{- \mathrm{det} \mathfrak G} \mathfrak G^{rt} = \sqrt{\mathrm{det} g} \frac{\rho}{h} \left( \frac{1}{a^2} - 1 \right) u^t |u^r| \ge 0. \]

The outer boundary term (at $r = r_\infty$) is more troublesome. If $r_\infty < \infty$, one has to assume that $\partial_t \delta \psi = 0$ and $\partial_r \delta \psi = 0$ at $r = r_\infty$ --- otherwise no result concerning stability can be established. However, if the above conditions are satisfied, then $dE_{(r_\ast,r_\infty)}/dt \le 0$, and the accretion flow is linearly stable.

For $r_\infty = \infty$ one has to inspect the asymptotic (for $r \to \infty$) behavior of $\delta \psi$. In our case, this is only possible for $\Lambda < 0$, and ``isothermal'' equations of state $p = ke$ with $k = 1/3, 1/2, 1$. The asymptotic behaviour of $\delta \psi$ is different in these three cases, and it is different from that described in~\cite{moncrief}. This issue is discussed below.

\section{Asymptotic behavior}

In what follows we assume $\Lambda < 0$ and consider only 3 cases: $p = ke$ with $k = 1/3, 1/2, 1$.

General bounds on the asymptotic falloff of $\partial_t \delta \psi$ and $\partial_r \delta \psi$ are given by requiring that the energy of perturbations $E_{(r_\ast,\infty)}$ is finite. They turn out to be marginally sufficient for the purpose of the stability proof, but the actual behavior of $\delta \psi$ --- controlled by Eq.~(\ref{aad}), governing the evolution of the perturbations --- is characterized by a faster falloff.

We restrict ourselves to perturbations whose asymptotic behavior can be characterized by
\begin{equation}
\label{bbb}
\delta \psi = \mathcal O \left(\frac{1}{r^n}\right), \quad \partial_t \delta \psi = \mathcal O \left(\frac{1}{r^n}\right), \quad \partial_r \delta \psi = \mathcal O \left(\frac{1}{r^{n+1}}\right),
\end{equation}
as $r \to \infty$, where $n > 0$ is a constant that should be determined. In other words, we assume that $\delta \psi$ must falloff at least as $1/r^n$. It is important that (\ref{bbb}) must hold for all times $t$ and all angles $\theta$ and $\phi$.

For $\Lambda = 0$ both $\sqrt{- \mathrm{det} \mathfrak G} \mathfrak G^{tt}$ and $\sqrt{- \mathrm{det} \mathfrak G} \mathfrak G^{rr}$ have the same asymptotic behaviour of $\mathcal O(r^2)$ (irrespectively of the assumed equation of state). The condition of the finiteness of $E_{(r_\ast,\infty)}$ implies --- assuming Eq.~(\ref{bbb}) --- that
\[ \partial_t \delta \psi = \mathcal O\left( \frac{1}{r^{3/2 + \epsilon}} \right), \quad \partial_r \delta \psi = \mathcal O \left( \frac{1}{r^{5/2 + \epsilon}} \right), \quad \mathrm{as} \quad r \to \infty,  \]
with $\epsilon > 0$. This suffices to show that $dE_{(r_\ast,r_\infty)}/dt \le 0$.

A careful reader can notice a difference between Eq.~(\ref{bbb}) and implicit assumptions made in \cite{moncrief}. For $\Lambda = 0$ it is enough to assume that the derivatives $\partial_t \delta \psi$ and $\partial_r \delta \psi$ have at most power-law falloffs $\partial_t \delta \psi = \mathcal O (1/r^n)$, $\partial_r \delta \psi = \mathcal O (1/r^{n^\prime})$, without specifying the relation between $n$ and $n^\prime$. Then, it follows form the condition $E_{(r_\ast,\infty)} < \infty$, that $\partial_t \delta \psi = \mathcal O (1/r^{3/2 + \epsilon})$ and $\partial_r \delta \psi = \mathcal O (1/r^{3/2 + \epsilon^\prime})$, where $\epsilon, \epsilon^\prime > 0$. This is sufficient to conclude that $d E_{(r_\ast,\infty)} /dt \le 0$, but such a reasoning cannot be repeated for $\Lambda \neq 0$ without providing additional information on $\partial_r \delta \psi$ and $\partial_t \delta \psi$.

\subsection{Asymptotic behavior for $p = e/3$}

For the equation of state $p = e/3$ one can show the following asymptotic behavior: $u^r = \mathcal O(r)$, $\rho = \mathcal O(1/r^3)$, $\rho/h = \mathcal O(1/r^2)$ (this follows straightforward from the explicit solutions given in Sec.~3). Thus, $\sqrt{- \mathrm{det} \mathfrak G} \mathfrak G^{rr} = \mathcal O(r^2)$, $\sqrt{- \mathrm{det} \mathfrak G} \mathfrak G^{tt} = \mathcal O(1/r^2)$, and $\sqrt{- \mathrm{det} \mathfrak G} \mathfrak G^{rt} = \mathcal O(r^0)$. The requirement of the finiteness of the energy $E_{(r_\ast,\infty)}$ yields, assuming Eq.~(\ref{bbb}), $\partial_r \delta \psi = \mathcal O(1/r^{3/2 + \epsilon})$, and $\partial_t \delta \psi = \mathcal O (1/r^{1/2 + \epsilon})$. This and Eq.~(\ref{bba}), allows to show that $d E_{(r_\ast,\infty)} /dt \le 0$. Indeed, we have $\sqrt{- \mathrm{det} \mathfrak G} \mathfrak G^{rt} (\partial_t \delta \psi)^2 = \mathcal O(1/r^{1+ 2\epsilon})$, and $\sqrt{- \mathrm{det} \mathfrak G} \mathfrak G^{rr} \partial_t \delta \psi \partial_r \delta \psi = \mathcal O( 1/r^{2 \epsilon})$.

A faster falloff of $\delta \psi$ can be inferred from Eq.~(\ref{aad}) (in addition to the requirement that $E_{(r_\ast,\infty)} < \infty$). In the case of the equation of state $p = e/3$, Eq.~(\ref{aad}) can be satisfied up to a leading order in the asymptotic expansion, provided that $\delta \psi = \mathcal O (1/r)$. These statements require an explanation. Equation (\ref{aad}) can be written as
\begin{eqnarray}
\label{ffg}
\mathfrak G^{tt} \partial_t^2 \delta \psi + 2 \mathfrak G^{tr} \partial_{tr}^2 \delta \psi + \frac{1}{\sqrt{- \mathrm{det} \mathfrak G}} \partial_r \left( \sqrt{- \mathrm{det} \mathfrak G} \mathfrak G^{rt} \right) \partial_t \delta \psi & & \\
+ \frac{1}{\sqrt{- \mathrm{det} \mathfrak G}} \partial_r \left( \sqrt{- \mathrm{det} \mathfrak G} \mathfrak G^{rr} \partial_r \delta \psi \right) + \frac{h a^2}{r^2 \rho} \Delta_\Omega \delta \psi & = & 0, \nonumber
\end{eqnarray}
where $\Delta_\Omega$ denotes the Laplacian on the 2-sphere. Assume now the asymptotic behavior of the solution in the form (\ref{bbb}). Then, taking into account the asymptotic falloff of the unperturbed solution for $p =e/3$, one can check that the leading order term in Eq.~(\ref{ffg}) is
\[ \frac{1}{\sqrt{- \mathrm{det} \mathfrak G}} \partial_r \left( \sqrt{- \mathrm{det} \mathfrak G} \mathfrak G^{rr} \partial_r \delta \psi \right) = (n-1) \mathcal O \left( r^{-n+2} \right). \]
Clearly, in order to satisfy Eq.~(\ref{ffg}) in the leading order, we have to demand that $n = 1$. This yields the claimed falloff $\delta \psi = \mathcal O (1/r)$. The above reasoning may by iterated, possibly yielding a stronger falloff condition. An analogous procedure can be also repeated for the cases with $p = e/2$ and $p = e$, that are discussed briefly in the following paragraphs.

\subsection{Asymptotic behavior for $p = e/2$}

In this case $\rho = \mathcal O(1/r^2)$, $\rho/h = \mathcal O(1/r)$, and $u^r = \mathcal O(r^0)$. Also $\sqrt{- \mathrm{det} \mathfrak G} \mathfrak G^{rr} = \mathcal O(r^3)$, $\sqrt{- \mathrm{det} \mathfrak G} \mathfrak G^{tt} = \mathcal O(1/r)$, and $\sqrt{- \mathrm{det} \mathfrak G} \mathfrak G^{rt} = \mathcal O(r^0)$. Here the situation is similar to that for $p = e/3$. Linear stability can be proved for perturbations satisfying $E_{(r_\ast,\infty)} < \infty$ and Eq.~(\ref{bbb}). Then $\partial_r \delta \psi = \mathcal O(1/r^{2 + \epsilon})$, $\partial_t \delta \psi = \mathcal O(1/r^{1 + \epsilon})$, and thus both terms appearing in Eq.~(\ref{bba}), $\sqrt{- \mathrm{det} \mathfrak G} \mathfrak G^{rt} (\partial_t \delta \psi)^2$ and $\sqrt{- \mathrm{det} \mathfrak G} \mathfrak G^{rr} \partial_t \delta \psi \partial_r \delta \psi$, behave asymptotically as $1/r^{2 + 2\epsilon}$ and $1/r^{2\epsilon}$, respectively.

Again, additional information provided by Eq.~(\ref{aad}) yield a faster falloff, namely $\delta \psi = \mathcal O(1/r^2)$. For the sake of brevity, we omit the details of this calculation.

\subsection{Asymptotic behavior for $p = e$}

In the case of ultra-hard equation of state $p = e$, Eq.~(\ref{aad}) is equivalent to a standard wave equation on the Schwarzschild--anti-de Sitter spacetime. A similar problem --- evolution of scalar fields in asymptotically Schwarzschild--anti-de Sitter spacetimes --- was investigated recently in~\cite{holzegel}.

For $p = e$, $\rho/h = \mathrm{const}$. We have $\sqrt{- \mathrm{det} \mathfrak G} \mathfrak G^{rr} = \mathcal O(r^4)$, $\sqrt{- \mathrm{det} \mathfrak G} \mathfrak G^{tt} = \mathcal O(r^0)$, $\mathfrak G^{tr} = 0$. Finiteness of the energy norm yields $\partial_r \delta \psi = \mathcal O(1/r^{5/2 + \epsilon})$, and thus $\partial_t \delta \psi = \mathcal O(1/r^{3/2 + \epsilon})$. The above bounds are again marginally sufficient for the stability --- one gets $\sqrt{- \mathrm{det} \mathfrak G} \mathfrak G^{rr} \partial_t \delta \psi \partial_r \delta \psi = \mathcal O(1/r^{2\epsilon})$.

The asymptotic behavior enforced by Eq.~(\ref{aad}) is $\delta \psi = \mathcal O(1/r^3)$.

\section{Summary}

We investigated conditions under which the Bondi-type accretion in Schwarzchild--(anti-)de Sitter spacetimes is stable. Stationary solutions were obtained in~\cite{kmm} for ``isothermal'' equations of state of the form $p = ke$, with $k = 1/3, 1/2, 1$. Polytropic solutions, corresponding to equations of state $p = K \rho^\Gamma$, were also investigated, although they cannot be written in a closed form. All these solutions appear linearly stable, provided that the perturbations vanish at the outer boundary of the cloud. If the accretion cloud can formally extend up to infinity --- this happens for the investigated ``isothermal'' equations of state and $\Lambda < 0$ --- one can show that the amplitude of perturbations with a finite energy decays sufficiently fast to ensure stability. 

One has to bear in mind that linear stability is only a prerequisite of (nonlinear) stability. Moreover, our analysis does not take into account the self-gravity of the accreting fluid. In principle, self-gravity of perturbations could play a role in the stability analysis even in cases when the self-gravity of the stationary, unperturbed flow can be safely neglected. However, a truly consistent treatment would require taking into account self-gravity of the unperturbed flow as well. Stationary solutions of self-gravitating polytropic fluid in asymptotically Schwarzchild--(anti-)de Sitter spacetimes were obtained numerically in~\cite{km}. Nonlinear stability of self-gravitating Bondi-type accretion flows without cosmological constant was investigated, also numerically, in~\cite{mm}.

\section*{Acknowledgments}

We would like to thank Piotr Bizo\'{n} for discussions.

This work has been partially supported by the following grants: Polish Ministry of Science and Higher Education grants IP2012~000172 (Iuventus Plus), 7150/E-338/M/2013, and the NCN grant DEC-2012/06/A/ST2/00397.

\end{document}